\documentclass{article}
\usepackage{ijcai23}

\usepackage{times}
\usepackage{soul}
\usepackage[hidelinks]{hyperref}
\usepackage{hyperref}
\usepackage[utf8]{inputenc}
\usepackage[small]{caption}
\usepackage{graphicx}
\usepackage{amsmath}
\usepackage{amsthm}
\usepackage{booktabs}
\usepackage{algorithm}
\usepackage[switch]{lineno}

\usepackage[T1]{fontenc}
\usepackage{amsfonts}
\usepackage{nicefrac}
\usepackage{microtype}
\usepackage{xcolor}

\usepackage{amssymb}
\usepackage{mathtools}
\usepackage{multirow}
\usepackage{import}
\usepackage{pifont}
\usepackage{textcomp}
\usepackage{color, colortbl}
\usepackage{subfigure}
\usepackage{makecell}
\usepackage{listings}
\usepackage{footmisc}
\usepackage[noend]{algpseudocode}
\usepackage{bm}
\usepackage[flushleft]{threeparttable}

\theoremstyle{definition}

\title{FusionAI: Decentralized Training and Deploying LLMs with Massive Consumer-Level GPUs}

\author{
Zhenheng Tang$^1$\and
Yuxin Wang$^1$\and
Xin He$^1$\and
Longteng Zhang$^1$\and
Xinglin Pan$^1$\and
Qiang Wang$^2$\and
Rongfei Zeng$^3$\and
Kaiyong Zhao$^6$\and
Shaohuai Shi$^2$\and
Bingsheng He$^4$\and
Xiaowen Chu$^5$\thanks{Correspondence to: Shaohuai Shi and Xiaowen Chu. }
\affiliations
$^1$Department of Computer Science, Hong Kong Baptist University\\
$^2$Harbin Institute of Technology, Shenzhen, $^3$Northeastern University\\
$^4$National University of Singapore, $^6$XGRIDS, $^5$Data Science and Analytics Thrust, HKUST(GZ)
\emails
\{zhtang,yxwang,csxinhe,ltzhang,csxlpan\}@comp.hkbu.edu.hk,\\
\{qiang.wang,shaohuais\}@hit.edu.cn, zengrf@swc.neu.edu.cn, \\
kyzhao@xgrids.com,
hebs@comp.nus.edu.sg, xwchu@ust.hk
}

\begin{document}

\maketitle

\begin{abstract}
The rapid growth of memory and computation requirements of large language models (LLMs) has outpaced the development of hardware, hindering people who lack large-scale high-end GPUs from training or deploying LLMs. However, consumer-level GPUs, which constitute a larger market share, are typically overlooked in LLM due to their weaker computing performance, smaller storage capacity, and lower communication bandwidth. Additionally, users may have privacy concerns when interacting with remote LLMs. In this paper, we envision a decentralized system unlocking the potential vast untapped consumer-level GPUs in pre-training, inference and fine-tuning of LLMs with privacy protection. However, this system faces critical challenges, including limited CPU and GPU memory, low network bandwidth, the variability of peer and device heterogeneity. To address these challenges, our system design incorporates: 1) a broker with backup pool to implement dynamic join and quit of computing providers; 2) task scheduling with hardware performance to improve system efficiency; 3) abstracting ML procedures into directed acyclic graphs (DAGs) to achieve model and task universality; 4) abstracting intermediate represention and execution planes to ensure compatibility of various devices and deep learning (DL) frameworks. Our performance analysis demonstrates that 50 RTX 3080 GPUs can achieve throughputs comparable to those of 4 H100 GPUs, which are significantly more expensive.\footnote{This work has been published on the Symposium on Large Language Models (LLM-IJCAI workshop 2023)}

\end{abstract}

\section{Introduction}\label{sec:intro}
\begin{figure}[t]
\vspace{-0.2cm}
    \subfigtopskip=2pt
    \setlength{\belowdisplayskip}{2pt}
    \setlength{\abovedisplayskip}{-5pt}
    \subfigbottomskip=2pt
    \subfigcapskip=1pt
   \centering
    {\includegraphics[width=0.8\linewidth]{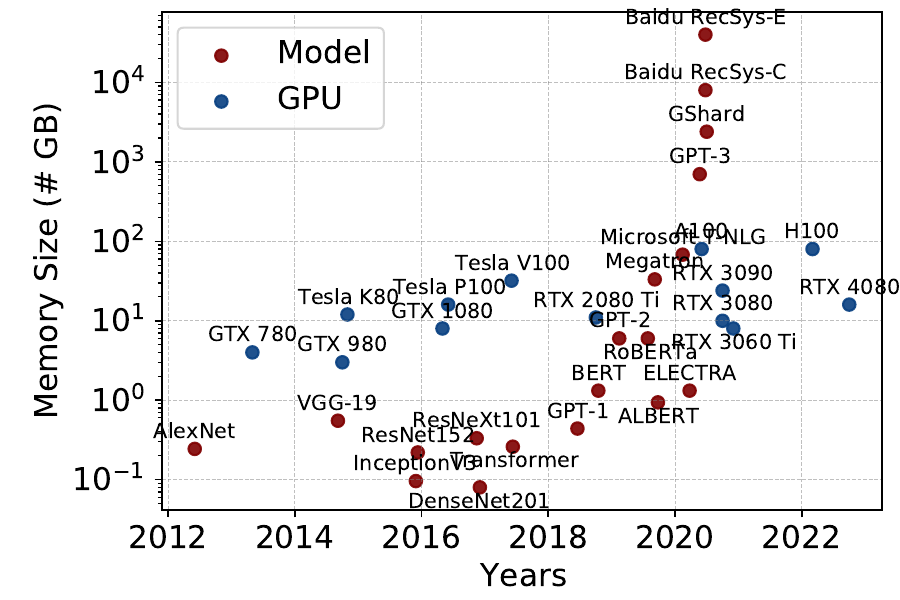}}
    \vspace{-0.3cm}
    \caption{The trends of GPU and model memory.}
    \label{fig:Mot_memory}
\vspace{-0.2cm}
\end{figure}

Deep neural networks (DNNs), from AlexNet~\cite{alexnet} to recent ChatGPT~\cite{chatgpt,scienceChatGPT}, have demonstrated significant benefits from larger training datasets and more parameters following scaling laws~\cite{ghorbaniscaling,alabdulmohsin2022revisiting,tay2022scaling}. However, the development of hardware has not kept pace with the rapidly increasing memory and computing power requirements, as shown in Figure~\ref{fig:Mot_memory}. As a result, current or future large language models (LLMs) require multiple high-end GPUs such as NVIDIA A100 and V100 for training and deployment, creating obstacles for researchers and engineers from smaller organizations who lack access to such resources.

Consumer-level GPUs, in contrast to high-end GPUs, are typically overlooked in LLMs due to their weaker computing performance, smaller storage capacity, and lower communication bandwidth. However, consumer-level GPUs are widely used in applications such as gaming, video editing, and data visualization and are sold in much larger quantities than high-end GPUs~\cite{NVIDIA2022}. Table~\ref{tab:GPUPerf} demonstrates that consumer-level GPUs possess impressive computing power, with comparable memory and FLOPS (floating-point operations per second) to high-end GPUs.

\begin{table}[t!]
\caption{Comparing different GPUs.}
\vspace{-0.5cm}
\begin{center}
\begin{small}
\resizebox{\linewidth}{!}{
\begin{tabular}{ccccc}
\toprule
\multirow{2}{*}{GPU} & TFLOPS & TFLOPS  & \multirow{2}{*}{Memory} & \multirow{2}{*}{Level} \\
&  (FP32) & FP32 Tensor Core &  \\
\bottomrule
RTX 4090 &  82.58 & 82.58  & 24GB & Consumer \\
RTX 4080 &  48.74 & 97.5 & 16GB & Consumer \\
RTX 3080 & 29.77 & 59.5 & 10GB & Consumer \\
H100 & 51.22 & 756 & 80GB & Data Center \\
A100 & 19.49 & 155.92 & 80GB & Data Center \\
\bottomrule
\end{tabular}
}
\end{small}
\end{center}
\vskip -0.1in
\label{tab:GPUPerf}
\end{table}

In addition, data privacy receives more and more attention these years~\cite{european_commission_regulation_2016}. Sharing raw user data with large companies to obtain the LLM services may expose user privacy. Splitting LLMs and running a part of it on the user-side may offer privacy benefits. Motivated from the computing power and privacy protection, we raise a crucial practical question:

\emph{Is it possible to unleash the potential vast untapped consumer-level GPUs and protect data privacy by disecting LLMs across massive decentralized devices?}

However, compared with training LLMs in data centers equipped with high-ends GPUs, mobile and personal devices present several critical challenges: 
1) \textbf{Limited CPU and GPU memory} necessitates more fine-grained partitions of DNN, distributed storage of datasets, and better scheduling algorithms to improve efficiency. 
2) \textbf{Incompatible software and hardware} of geographically distributed devices may make programming complicated.
3) \textbf{The variability of collaboration} due to dynamic joins and exits significantly brings in new challenges of fault tolerance and rescheduling.
4) \textbf{Low network bandwidth} may lead to unacceptable communication times, especially with the substantial amount of data exchanged between devices.
5) \textbf{Hardware performance heterogeneity} in terms of various GPU and CPU architectures, memory sizes, bandwidth and battery size, necessitates consideration of diverse constraints.

To address these problems and unlock the vast computing power of consumer-level GPUs, we proposes a decentralized computing system equipped with following system designs (Section~\ref{sec:design}):
\begin{itemize}
    \item Dissecting directed acyclic graphs (DAGs) of model execution into sub-DAGs and loading them onto devices with limited memory~\cite{EIPaving};
    \item Abstracting operations of DAGs into the intermediate representation (IR) plane and the execution plane to implement compatibility of various hardware and software.
    \item Pooling a part of computing providers as backup nodes for fault tolerance.
    \item Designing partitioning and pipeline strategies to reduce communication costs.
    \item Predicting computation and communication costs and scheduling tasks on devices to tackle hardware performance heterogeneity.
\end{itemize}




\section{Related Works}\label{sec:related}
\subsection{Centralized Distributed DL}\label{sec:Dist}

Traditional training and inference of large DL models are distributed within a single organization via high-speed local area network (LAN). In \textbf{Data parallelism (DP)}, input data is divided into subsets and processed on duplicated models across different machines, with the new gradients or parameters from these models being aggregated to update the original model~\cite{tang2020communication,shi2019distributed}. However, the scalability of DP is limited by the inefficiency of large-batch SGD~\cite{mccandlish2018empirical,9275615}, high communication costs of the whole model size~\cite{tang2020communication,9155269,9275615,shi2019convergence}, and the inability to load an LLM onto a single GPU.
\textbf{Pipeline parallelism (PP)} dissects model layers into multiple stages and executes them sequentially on different devices, communicating intermediate results between stages~\cite{huang2019gpipe}. Improving the pipelines between execution and communication can reduce bubble time and improve overall efficiency~\cite{narayanan2019pipedream}. However, the layer-wise dependency of forward and backward processes limits the scalability of PP~\cite{park2020hetpipe}.
\textbf{Tensor parallelism (TP)} vertically splits model stages across multiple devices, with each group aggregating computing results and sending them to the other group for the next model stage. TP is suitable for super-large models in high communication-bandwidth environments~\cite{narayanan2021efficient}, due to its more fine-grained computation and communication overlaps.

Most of current advanced and high-efficient training systems~\cite{fairscale,sergeev2018horovod,byteps,shoeybi2019megatron,rasley2020deepspeed} focus on training in data center, instead of decentralized training.

\subsection{Decentralized DL}\label{sec:opentraining}
Peer-to-peer computing~\cite{DONet,chu2014dissecting,diskin2021distributed} and scientific computing~\cite{5719609} have previously explored the potential of utilizing massive personal and edge devices. Walle~\cite{Walle} is a device-cloud ML system that distributes ML tasks with small models between devices, while MOSAIC~\cite{mosaic} and JALAD~\cite{8645013} focus on efficient inference and only verify their systems on small models and datasets. ComAI~\cite{9796769} proposes a cloud-edge collaborative inference framework.

Although many current edge works focus on efficient inference, federated or transfer learning with multiple small models~\cite{EIPaving,9628184,9355592,tang2022gossipfl,tang2022virtual}, our system concentrates on exploiting massive devices for both training and inference of LLMs. DeDLOC~\cite{DeDLOC} implements volunteer computing to train ALBERT~\cite{Lan2020ALBERT} with DP because ALBERT is small enough to be loaded onto a single V100.

SWARM~\cite{ryabinin2023swarm}, Learning@home~\cite{DecentMOE}, CocktailSGD~\cite{cocktailSGD}, AQ-SGD~\cite{NEURIPS2022_7a43b8eb} and DT-FM~\cite{yuandecentralized} are the first to use low-bandwidth connected devices to train larger models such as GPT-2 and large Transformers. However, they do not consider the heterogeneity of consumer-level GPUs, the variability of collaboration, the compatibility of software and hardware, or the generality of DL tasks.




\subsection{Communication Efficiency}\label{sec:communicationcomp}
Communication compression techniques play a crucial role in addressing the communication challenges faced by distributed deep learning systems. 

Sparsification~\cite{EffiUseofLimitMemory,li2022on,shi2021towards,zhang2023evaluation} means to pick up a part of values in the gradients or model parameters to communication.

Quantization~\cite{ProbroundingNN,DorefaNet,QSGD,TernGrad} uses lower bits to represent data originally represented by 32 bits on each dimension of the transmitted gradient.

Local-SGD~\cite{NIPS2019_9288,Yu2018ParallelRS,spiridonoff2021communicationefficient} permits flexible communication frequencies. Specifically, each worker independently executes several or more iterations before averaging all local models to obtain the most recent global model. 

The asynchronous parallel SGD framework allows the server to update the global model with the updates from a part of workers instead of all workers~\cite{hogwild,NEURIPS2022_029df12a}. Thus, it enables more independent updates of the workers and reduces one-round data transmission during communication.

To reduce the large communication time caused by the extremely low communication bandwidth and the enormous communication data during training LLMs, the FusionAI incorporates these techniques and conduct scheduling with them.

\subsection{Edge DL}
Edge devices have limited resources in terms of memory, computational power, communication, and energy, which encourages memory-efficient and data-efficient training and inference. Low-precision training~\cite{NEURIPS2018_335d3d1c,cambier2020shifted} and trading memory for computation~\cite{gruslys2016memoryefficient,chen2016training} are designed for high-throughput cloud training of large models with high storage requirements, but may not be suitable for low-storage edge devices. This work aims to provide scalability for large-scale dataset and model training and inference on edge devices by partitioning the workload of large model training into multiple subtasks, adapting to different device topologies.


\subsection{Incentive Mechanism}\label{sec:incentive}


Incentive mechanisms serve as economic catalysts to facilitate decentralized training. Some previous schemes like auction, contract, Stackelberg game, etc. can be employed to incentivize some selfish and rational clients to participate in this cooperative learning~\cite{zeng2021comprehensive,zeng2020fmore}. Compared with previous incentive schemes in other similar scenarios, there are some special challenges and considerations in the mechanisms design. (1) The property of online training indicates that the arrival and departure time is unknown and varies drastically for different clients. This online property is in line with asynchronous training but renders previous one-round incentive schemes ineffective. (2) Some other alternative choices should be considered for clients like bitcoin mining, client-assisted contributions, etc.,  which can also bring some economical benefits to their participation. In other words, incentive design should attract clients to join in decentralized learning from many candidate choices. (3) The incentive scheme should be robust and resilient to some malicious clients which contribute nothing but endeavor to get large paybacks. 
 

\subsection{Security}\label{sec:security}

The security challenges in decentralized computing are also a critical concern. The potential risks of infringement privacy are the unauthorized data access and manipulation~\cite{hitaj2017deep,zhu2019deep}. There is also the threat of malicious computing nodes that can disrupt system operations and compromise model robustness~\cite{bagdasaryan2020backdoor}. The issue of membership inference can expose if a specific data point was used in training the model~\cite{nasr2019comprehensive}. In response to these challenges, privacy-preserving techniques such as secure multiparty computation, differential privacy, and encryption~\cite{araki2016high,geyer2017differentially} are suggested as mitigation strategies.

\subsection{Hardware Performance Modeling}\label{sec:PerfModel}
Works that predict computation and communication time can be divided into the DNN-based and analytic performance model. DNN-based methods~\cite{8622396} train a DL model to predict the execution time for a given model. In our scenario, DNN-based models impose a significant computational burden on low-computing-ability nodes due to the large overhead required for training and storage. More crucially, computing providers are dynamic and heterogeneous, making it difficult to adapt the learned performance model to the newly joined nodes. The analytic performance model estimates the hardware performance based on model sizes, communication bandwidth and latency, and FLOPS of GPUs and CPUs. With a short period of profiling to fit a few parameters, analytic performance models~\cite{qi2017paleo,park2020hetpipe} can approximate the hardware performance with small errors.

\subsection{Energy and Carbon Footprint}\label{sec:energy}





Currently, the energy consumption of high-end GPUs has become a bottleneck for training large models~\cite{tang_dl2019,zeus_you_nsdi2023,dynamic_wang_tpds2022}. In contrast, our proposd FusionAI can address this bottleneck by providing feasibility in terms of power consumption and offering more optimization opportunities and flexibility for energy efficiency. By leveraging energy optimization via power management, adaptive batching and alignment and scheduling~\cite{batch_dvfs_tpds2022,energy_wang_tgcn2022}, further improvements can be made to enhance global performance and energy efficiency.

\section{System Design}\label{sec:design}
\begin{figure}[h]
\vspace{-0.5cm}
    \subfigtopskip=2pt
    \setlength{\belowdisplayskip}{2pt}
    \setlength{\abovedisplayskip}{-5pt}
    \subfigbottomskip=2pt
    \subfigcapskip=1pt
   \centering
    {\includegraphics[width=0.99\linewidth]{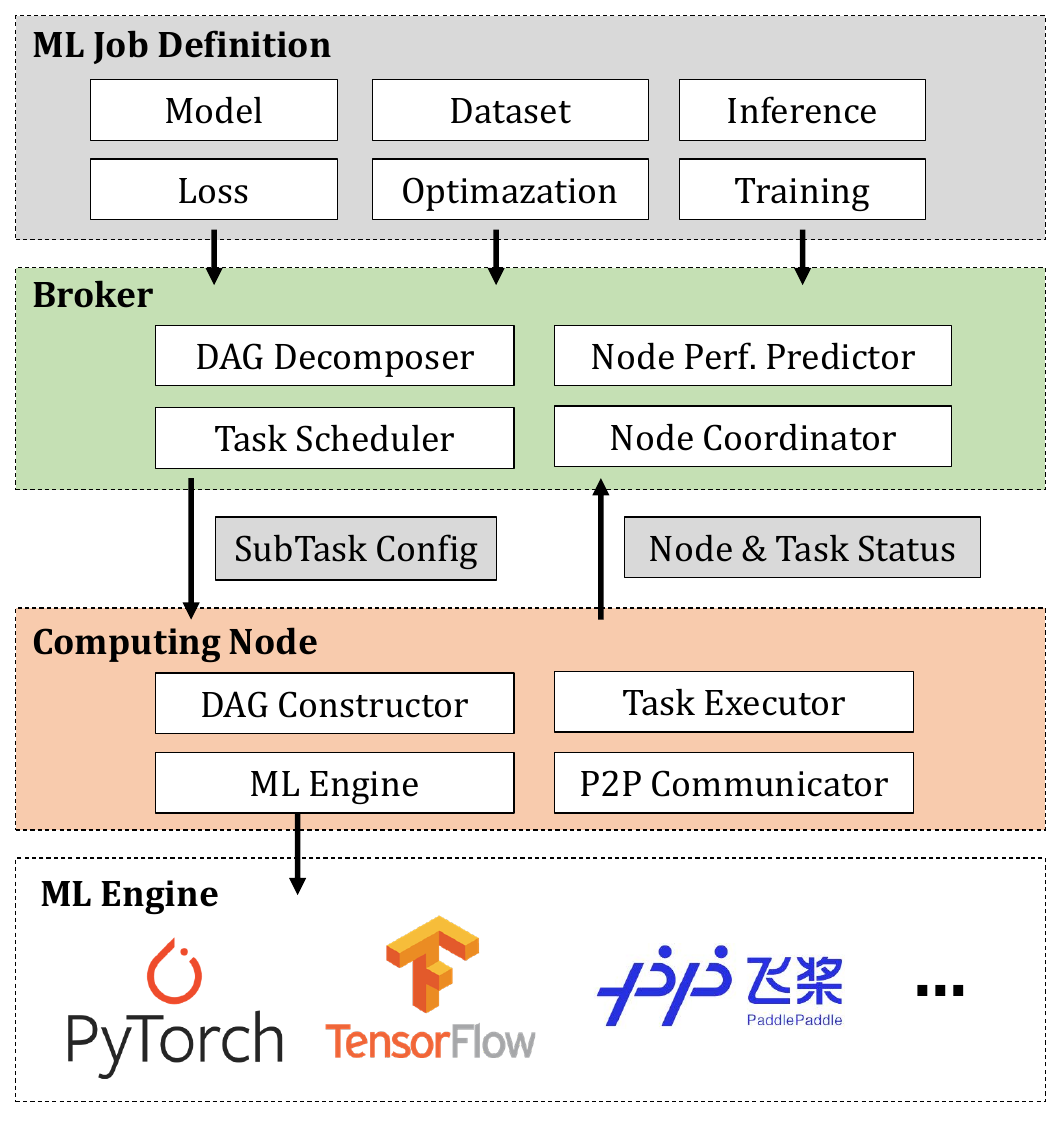}}
    \vspace{-0.3cm}
    \caption{The overview of system design of FusionAI.}
    \label{fig:FrameworkFigure}
\vspace{-0.3cm}
\end{figure}

We firstly give an overview of design goals and the high-level framework (Figure~\ref{fig:FrameworkFigure}) of our system in this section. Subsequently, we delve into components presented in Figure~\ref{fig:FrameworkFigure} in the subsequent sections.

\subsection{Overview}\label{sec:overview}

The aforementioned unique properties render the direct application of traditional distributed training and inference algorithms infeasible for our system.  Thus, in our system, we envision a novel system design that enables the utilization of consumer-level GPUs connected via the Internet for computing large models. Unlike traditional distributed training and inference systems, our system aims to simultaneously achieve the following goals:
\begin{itemize}
\item \textbf{P1 User autonomy and robustness}: This system allows users with computational devices to autonomously join or leave as computing nodes (Compnode). New nodes can join at any time, while existing nodes can exit due to various reasons.
\item \textbf{P2 High scalability}: This system supports a large number of computing nodes to participate, resulting in a high acceleration ratio of system performance.
\item \textbf{P3 Device compatibility}: This system does not impose restrictions on user device performance. Users can have highly heterogeneous GPUs, CPUs, and network bandwidths, ranging from mobile GPUs to mid-range GPUs in PCs or high-performance GPUs. Network performance, CPUs, storage, and other resources also exhibit high heterogeneity.
\item \textbf{P4 DL framework compatibility}: This system does not limit the choice of DL computing frameworks for participating nodes. Users can utilize PyTorch~\cite{torch}, TensorFlow~\cite{abadi2016tensorflow}, PaddlePaddle~\cite{paddle}, or other deep learning frameworks
\item \textbf{P5 Model universality}: This system provides support for a wide range of and customizable  model architectures.
\item \textbf{P6 Task universality}: This system accommodates various loss functions, such as unsupervised loss and contrastive learning loss, as well as various training tasks, including image classification, regression tasks, and NLP language understanding tasks.
\end{itemize}



To achieve \textbf{P1}, we abstract each computing provider as a compnode, and a compnode manager as a broker. When compnodes join the collaboration, the broker registers its basic information. When a compnode quits for various reasons, the broker should be aware and schedule the unfinished tasks to other compnodes.

For \textbf{P2}, the broker should schedule and assign tasks with suitable workloads to compnodes based on their hardware performance. The scheduling and training algorithm should identify the system bottleneck and reduce it to increase system throughputs.

For \textbf{P3},\textbf{P4}, the most important is to disassemble the dependency between hardware and DL frameworks, model definition, To this end, We first abstract different execution processes of a DL model into Forward Propagation (FP), Backward Propagation (BP) and Update. The inference of a model is the FP process, and obtaining gradients needs FP and BP processes, conducting SGD update needs FP, BP and Update. The FP and BP processes are formulated as the directed acyclic graph (DAG). Nodes in DAG represent the computing operators, like addition, matrix multiplications, convolution operation, cross entropy and so on. Directed edges represent the message passing between nodes, meaning that the outputs of the source node will serve as inputs to the target node. Second, we abstract the DAG into the \textit{intermediate representation (IR) plane} and the \textit{execution plane}. Users only need to provide the definition of the DAG in the IR plane. Then, the original complete DAG can be decomposed into sub-DAGs to be reconstructed and executed on different compnodes according to the scheduling. The execution plane is responsible for designing and implementing general interfaces to adapt different ML Engines to execute DAGs. Thus, the compnodes can utilize devices and DL frameworks according to their preference. 

To achieve \textbf{P5} and \textbf{P6}, DAG definition should allow users to customize their model definitions and DL tasks. We design a unified interface for new DAG operators to allow them can be automatically incorporated in the IR and execution planes. Thus, not only the classification tasks, but also some advanced contrastive learning~\cite{chen2020simple} and semi-supervised learning~\cite{van2020survey} tasks can be inplemented in our system.

\subsection{Broker}\label{sec:broker}
As introduced in Section~\ref{sec:overview}, the broker bridges job submitters (users) and compnodes (Section~\ref{sec:compnode}) together. Computing provider can join the system and register as a compnode. Each compnode is assigned with a unique ID. And the broker is responsible for periodically sending the ping-pong signal for detecting whether the compnode is offline or not. The broker periodically sends ping-pong signals to detect the online status of compnodes. Alternatively, compnodes can perform ping-pong tests among their dependent peers to determine their online availability. In the event that a compnode is offline and has unfinished tasks, the broker selects a replacement from the backup compnode pool.


Upon receiving new jobs from users, the broker processes the job definition file, which contains customized models, through the DAG decomposer, resulting in sub-tasks (Section~\ref{sec:task-split}). During task scheduling, the broker utilizes the hardware performance predictor (Section~\ref{sec:model-hardware}) to assign tasks to compnodes with balanced workloads (Section~\ref{sec:task-schedule}).

\subsection{Computing Node}\label{sec:compnode}
The decentralized computing system can be represented as a bidirectional graph $\mathcal{P}=\langle \{p^{i}\}_{i=1}^{n_p}, \{(p^{i}, p^{j})_{1\leq i < j\leq n_p}\}\rangle$ with $n_p$ nodes. Each peer (node $p$) is equipped with diverse GPUs, CPUs and disk storage denoted as $D_{gpu}^p$, $D_{cpu}^p$ and $D_{disk}^p$, respectively. The computation speed $S(p)$ of a compnode is measured by FLOPS (floating-point operation per second). The communication cost between peers $p^{i}$ and $p^{j}$ can be characterized by the alpha-beta model~\cite{MPICH,shi2019mg}: $T_{comm}^{ij}(M)=\alpha^{ij} + \beta^{ij}M$, where $\alpha^{ij}$ is the latency component, $\beta^{ij}$ the inverse of the communication speed, and $M$ the message size.


Compnodes exhibit variations in hardware performance and collaboration periods, and can be categorized into supernode and antnode.  \textbf{Supernodes} provide long-term and stable computing, storage and communication services, while \textbf{Antnodes} dynamically participate in collaboration with relatively lower computing power, storage, and communication bandwidth.

Upon receiving the assigned tasks with the IR of the sub-graph, compnodes construct the sub-DAGs and execute them (Section~\ref{sec:task-split}), facilitating communication of outputs via the decentralized communicator as message passing (Section~\ref{sec:decentralized-comm}) within the original DAG.




\subsection{Decentralized Communication}\label{sec:decentralized-comm}
In our system, various types of data require communication among compnodes, including training and testing raw data, hidden features, model weights, gradients, etc. Unlike traditional distributed training approaches that rely on a central parameter server~\cite{tang2020communication}, our system adopts a decentralized communication topology. 


For efficient management of distributed storage and lookup of data, we leverage the power of Distributed Hash Table (DHT)~\cite{ratnasamy2001scalable}. DHT enables the distribution of data across the network by employing key-value pairs and hashing, allowing for rapid and efficient retrieval based on unique keys. By utilizing DHT, our system achieves a decentralized and self-organizing architecture. Each compnode independently stores and retrieves data, making the system resilient to individual node failures. The use of DHT enhances scalability, fault tolerance, and flexibility in our decentralized communication framework.



\begin{figure}[h]
\vspace{-0.1cm}
    \subfigtopskip=2pt
    \setlength{\belowdisplayskip}{2pt}
    \setlength{\abovedisplayskip}{-5pt}
    \subfigbottomskip=2pt
    \subfigcapskip=1pt
   \centering
    {\includegraphics[width=0.99\linewidth]{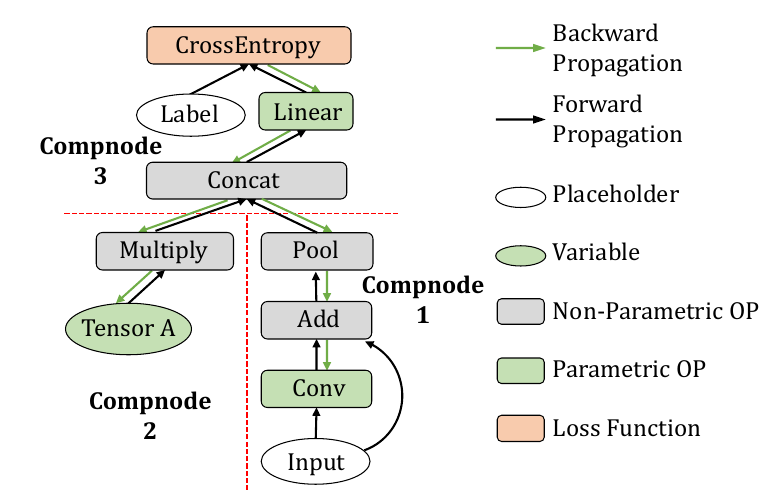}}
    \vspace{-0.1cm}
    \caption{DAG of FP and BP processes.}
    \label{fig:DAG}
\vspace{-0.1cm}
\end{figure}

\begin{table*}[ht!]
\caption{OP nodes and their attributes in the example DAG (Figure~\ref{fig:DAG}).}
\vspace{-0.5cm}
\begin{center}
\begin{small}
\begin{tabular}{ccccccc}
\toprule
OP names & OP users & Type & Args & Kwargs & Compnode location & Compnode users \\
\bottomrule
Input & Conv, Add & Placeholder & -  & - & 1 & 1 \\
Conv & Add & Parametric OP & Input & - & 1 & 1 \\
Add & Pool, Multiply & Non-Parametric OP & Conv & - & 1 & 1, 2 \\
Pool & Concat & Non-Parametric OP & Add & - & 1 & 3 \\
Tensor A & Multiply & Variable & -  & - & 2 & 2 \\
Multiply & Concat & Non-Parametric OP & Tensor A & - & 2 & 3 \\
Concat & Linear & Non-Parametric OP & Multiply, Pool & - & 3 & 3 \\
Linear & CrossEntropy & Parametric OP & Concat & - & 3 & 3 \\
Label & CrossEntropy & Placeholder & -  & - & 3 & 3 \\
CrossEntropy & - & Loss Function & Label, Linear & weight: 1.0 & 3 & 3 \\
\bottomrule
\end{tabular}
\end{small}
\end{center}
\vskip -0.1in
\label{tab:OPNodes}
\end{table*}

\begin{table*}[ht!]
\caption{Sub-graphs and its attributes in the example DAG (Figure~\ref{fig:DAG}).}
\vspace{-0.5cm}
\begin{center}
\begin{small}
\begin{tabular}{ccccccc}
\toprule
Subgraph & Compnode & Nodes & Inner required data & Outer required data & Outwards data & Compnode users \\
\bottomrule
\multirow{2}{*}{1} & \multirow{2}{*}{1} & Input, Conv,  & Input, Conv,  & Input & Add, Pool, & \multirow{2}{*}{2,3} \\
&  & Add, Pool   &   Add  &  &  & \\
\midrule
\multirow{2}{*}{2} & \multirow{2}{*}{2} & Tensor A,  & Tesor A & Add & Multiply & \multirow{2}{*}{3} \\
&  &  Multiply   &    &  &  & \\
\midrule
\multirow{2}{*}{3} & \multirow{2}{*}{3} & Concat, Linear, & Label, Concat, & Label, Pool,  & - & - \\
&  &   Label, CrossEntropy  &  Linear &  Multiply &  & \\
\bottomrule
\end{tabular}
\end{small}
\end{center}
\vskip -0.1in
\label{tab:subgraphs}
\end{table*}

\subsection{Sub-Tasks: FP, BP and Optimization}\label{sec:task-split}

The FP and BP processes of neural networks can be formalized as a directed acyclit graph $\mathcal{G} = \langle \{ o^{i}\}_{i=1}^{n_o}, \{(o^{i},o^{j})\}  \rangle$, where $n_o$ is the total number of operators, node $o^{i}$ represents an operation $f^{i}$. The directed edge $(o^{i},o^{j})$ indicates that $o^{j}$ cannot start until $o^{i}$ has finished, and the outputs of $o^{i}$ will be sent to $o^{j}$.

Figure~\ref{fig:DAG} illustrates one of the DAGs for the FP and BP processes. The operators (OPs) are categorized into parametric and non-parametric ones. Parametric OPs have parameters that require gradients for optimization, while non-parametric OPs do not. The leaf nodes include placeholders, which do not require gradients (e.g., inputs and labels), and variables that need to be optimized (e.g., malicious samples in adversarial attacks~\cite{adv_attack} or style-control variables in StyleGAN~\cite{karras2019style}). During the forward pass, all leaf nodes and OPs participate in the calculations. In the backward pass, placeholders do not require backward computation. During model optimization, non-parametric OPs and placeholders are not optimized, and their parameters do not need to be communicated. However, the parameters of parametric OPs and variables require to be optimized and synchronized with the supernode in case of compnode failures.


In Figure~\ref{fig:DAG}, the OPs are assigned to different compnodes for execution. The outputs of preceding OPs are used as inputs to subsequent OPs to which they are directed. If OPs are assigned to the same compnode, the data passing between them is local and fast, as indicated by the gray lines. However, data passing between OPs in different compnodes consumes communication resources, as shown by the black lines. The inputs and labels can be retrieved from the data provider through the Distributed Hash Table (DHT), as described in Section~\ref{sec:dataset-storage}.



Each sub-task for a compnode is defined by its task type and sub-graph, represented as $\mathcal{G}_{\mathcal{S}_k} = \langle \{ o^{i}\}_{i\in \mathcal{S}_k}, { o^{i},o^{j} } \rangle$, where $\mathcal{S}_k$ is the set of node indexes in the sub-graph of task $k$. Figure~\ref{fig:DAG} shows that the entire DAG is partitioned into three parts and assigned to compnodes $1$ to $3$. Upon receiving task configurations, compnodes reconstruct these sub-DAGs, load or initialize the parameters of the parametric OPs, and execute the assigned sub-tasks.

\subsection{Executing Sub-Tasks}\label{sec:task-execute}
We employ a task executor to manage the message passing between OPs and perform the computations of the OPs with their inputs. Table~\ref{tab:OPNodes} provides an overview of all OP nodes in the DAG~\ref{fig:DAG} and their attributes. The OP Users and Args are defined based on the FP process and can also be utilized in the BP process. Their explanations are provided with respect to the FP task, followed by BP and Update tasks.

\textbf{FP task.} The FP edges in DAG (Figure~\ref{fig:DAG}) can be identified by the OP names and the OP users. For example, the FP edge between the \textbf{FP Task:} The FP edges in the DAG (Figure~\ref{fig:DAG}) can be identified by the OP names and the OP users. For example, the FP edge between the Input and Conv OPs is determined by that the Conv OP is the user of the Input OP, and the edge is directed towards the Conv OP. The Args of an OP specify data from which OPs are required. The Kwargs identify constant values. When executing, each sub-graph finds all the necessary inputs from other sub-graphs. Once all the inputs are available, the FP task is launched and executed. The outputs of an OP are passed to its OP Users. When the executor detects that the OP Users are located on an external compnode, it invokes communication functions to send the outputs to the corresponding external compnodes. For example, the executor on compnode 1 will transmit the outputs of the Add OP to the Multiply OP on compnode 2.


\textbf{BP task.} In most cases, the BP edges are the reverse of the FP edges, except for the edges directed towards leaf nodes that do not require gradients, such as the Input and Label placeholders. The required resources for BP tasks change to gradients from OP Users, instead of data from OPs in Args. Similar to the FP task, the computed gradients are returned to their Arg Nodes. If the node is located on an external compnode, the gradients are also communicated to that compnode. For example, the executor will send the gradients on the outputs of the Pool OP from compnode 3 to compnode 1 (since these gradients are computed by the Concat OP on compnode 3).

\textbf{Update task.} To support adaptive optimizers for different parametric OPs, users can define optimizers and corresponding hyperparameters in the configuration file. The broker assigns the appropriate optimizers to the target compnode based on its assigned OPs. Once the BP tasks are completed, i.e., the gradients are computed for the parametric OPs, the executor can utilize these optimizers to begin the optimization process.

\textbf{Message passing.} The executor possesses all the information about the sub-graph and can control the data flow between OP nodes. However, the executor does not have direct control over external OP nodes, which prevents it from moving data to those nodes. Therefore, for message passing between compnodes, the send-side executor must determine which compnode should receive the data, and the receive-side executor registers the message processing functions on the compnode to store and process the required data. Table~\ref{tab:subgraphs} presents the attributes of subgraphs on compnodes and their associated attributes, which are utilized for efficient and convenient message passing. The outputs of certain OPs can be quickly evaluated to determine whether they should be sent out or kept locally. When a compnode receives data, it can be stored in the executor if it is part of the Outer Required Data, and vice versa.

\subsection{Modeling Hardware Performance}\label{sec:model-hardware}

In this section, we introduce how to record and measure the hardware performance of a compnode. As mentioned in Section~\ref{sec:compnode}, compnodes provide information about their GPU ($D_{gpu}^p$), CPU ($D_{cpu}^p$), and disk storage ($D_{disk}^p$). However, the actual computation speed ($S(p)$) may not reach the peak performance ($S^*(p)$) specified by the peers. This can be attributed to various factors, including the complexity of customized libraries like MKL~\cite{MKL} and cuDNN~\cite{chetlur2014cudnn}, as well as other parallel computations. To estimate the actual computation speed $S(p)$,  a regression-based scaling-down factor~\cite{qi2017paleo} $\lambda_p$ is used to estimate the actual computation speed $S(p)=S^*(p)\lambda_p$. Before scheduling, $\lambda_p$ is fitted by a short-time profiling. To model the computation time ($T(f, p)$) of a specific operation $f$ on peer $p$, we incorporate the computation performance model of PALEO~\cite{qi2017paleo} into our work:
\begin{equation}
    T(f, p) = \mathcal{R}(\text{Pa} (f)) + \mathcal{C}(f,p) + \mathcal{W}(f,p).
\end{equation}
The computation time $\mathcal{C}(f,p)$ is calculated as the FLOP (floating-point opration) counts of the operation $f$ divided by the $S(p)$: $\mathcal{C}(f, p)=\text{FLOPs}(f)/S(p)$. While the $\mathcal{R}(\text{Pa} (f))$ represents the time of retriving data from parent nodes of $f$, $\mathcal{W}(f,p)$ the time of writing data into memory. When $f$ and its parents are located on different compnodes, $\mathcal{R}(\text{Pa} (f))$ operation involves the communication and loading time between $f$ and $\text{Pa}(f)$. A DAG of the neural network has operators that can be executed sequentially or in parallel. As a result, the executing time $T(\mathcal{G}_{\mathcal{S}_k})$ of a sub-graph $\mathcal{G}_{\mathcal{S}_k}$ on peer $p$ is within the range $[\text{max}_i T(f^{i},p), \sum^{i} T(f^{i},p) ]_{i\in\mathcal{S}_k}$.


\subsection{Tasks Scheduling}\label{sec:task-schedule}
We consider training a LLM with a group of peers, where each peer is responsible for processing a sub-DAG of the LLM. Given the hardware performance obtained in Section~\ref{sec:model-hardware} and partitioned sub-DAGs $\{\mathcal{S}_k\}_{k\in\mathcal{K}}$ ($\mathcal{K}$ is the set of tasks), and a group of compnodes $\mathcal{P}$, we propose the task scheduling problem as below:
\begin{align}
\min_{\mathcal{A}} \max_{p\in{\mathcal{P}}}   \ \ & \sum_{k\in\mathcal{A}_p} T(\mathcal{G}_{\mathcal{S}_k})     \label{eq:schedule}  \\
s.t. \ \  &  D_{gpu}^p \geq  \sum_{k\in\mathcal{A}_p}  D_{gpu}(\mathcal{G}_{\mathcal{S}_k}),  \notag \\
 \ \  &  D_{cpu}^p \geq  \sum_{k\in\mathcal{A}_p} D_{cpu}(\mathcal{G}_{\mathcal{S}_k}),  \notag \\
 \ \  &  D_{disk}^p \geq  \sum_{k\in\mathcal{A}_p}  D_{disk}(\mathcal{G}_{\mathcal{S}_k}),  \notag
\end{align}
in which $D_{gpu}(\mathcal{G}_{\mathcal{S}_k})$, $D_{cpu}(\mathcal{G}_{\mathcal{S}_k})$ and $D_{disk}(\mathcal{G}_{\mathcal{S}_k})$ represent required GPU, CPU and disk memory to execute sub-DAG $\mathcal{G}_{\mathcal{S}_k}$,  $\mathcal{A}_p \in \mathcal{A}$ represents the tasks assigned to peer $p$. The solution of problem~\ref{eq:schedule} can generate the load-balanced sub-DAGs to peers.

\subsection{Datasets Storage and Distribution}\label{sec:dataset-storage}
Users who submit the ML jobs can choose the dataset sources. They can use public shared datasets or their own datasets to train their models. Similar with data flow between OPs, data providers also communicate input and labels through DHT as described in Section~\ref{sec:decentralized-comm}.

\textbf{Public datasets.} Because the supernodes are stable and consistent in collaboration, some commonly used datasets can be stored on them. By employing suitable incentive mechanisms, data providers can receive credits or benefits for sharing these datasets. During training, compnodes that have Input or Label placeholders consistently retrieve data from these data providers.

\textbf{Private datasets.} Users can provide the customized datasets by themselves. If privacy is not a concern, users can choose to upload their datasets to the super nodes and have the compnodes retrieve the data from there. However, if privacy is a concern, users can act as compnodes with operators near the input. During DAG execution, users only need to send intermediate features to other compnodes without sharing weights, which helps prevent model inversion attacks~\cite{wang2021variational}. Similarly, to ensure label privacy, users can compute the loss themselves by receiving outputs from other compnodes without sharing the labels.



\section{Analysis of System Performance}\label{sec:analysis}
In this section, we approximate the time of the FP of the DAG. Following Section~\ref{sec:task-schedule}, the computation time of each device is approximated by $T_p=\sum_{k\in\mathcal{A}_p} T(\mathcal{G}_{\mathcal{S}_k})$. Because the tensor parallelism would introduce much more extra communication costs between devices,  we simply consider the pipeline parallelism in this paper for now. Thus, for most famous deep models, sub-DAGs are sequentially executed. Because the read and write time in the local device is little comparing with the training and communication, we can remove $\mathcal{R}(\text{Pa}(f))$ if $\text{Pa}(f)$ and $\mathcal{W}(f,p)$ local on the same $p$ with $f$.

Then we can formalize the latency of FP process as:
\begin{small}
\begin{align}
   T(\mathcal{G})_{lat}  & = \sum_p^{p\in\mathcal{P}} T_p = \sum_p^{p\in\mathcal{P}} \sum_{k\in\mathcal{A}_p} T(\mathcal{G}_{\mathcal{S}_k})
   = \sum_p^{p\in\mathcal{P}} (\mathcal{C}_p + \mathcal{R}_p) \notag \\ 
    \mathcal{C}_p  & = \sum_{k\in\mathcal{A}_p} \sum_{f \in\mathcal{S}_k} \mathcal{C}(f,p) \notag \\
    \mathcal{R}_p  & = \sum_{k\in\mathcal{A}_p} \sum_{ \substack{ f \in\mathcal{S}_k, \\ P(f) \neq P(\text{Pa}(f))}}  \mathcal{R}(\text{Pa}(f)),
    \label{eq:time_latency}
\end{align}
\end{small}
in which $P(f)$ maps operator $f$ to its located peer $p$, $M_f$ represents the size of outputs from operator $f$, $\mathcal{R}(\text{Pa}(f)))= T_{comm}^{P(f,\text(Pa)(f)) }(M_f)$. When pipelining processes of different batches during training, some computation time and communication time can be overlapped. Considering $n_b$ batches are pipelined, the time cost of processing them is:
\begin{small}
\begin{equation}\label{eq:time_pipe}
   T(\mathcal{G})_{n_b,pipe}  = \sum_p^{p\in\mathcal{P}} (\mathcal{C}_p + \mathcal{R}_p) + (n_b - 1) \max_{p\in\mathcal{P}} (\mathcal{C}_p, \mathcal{R}_p).
\end{equation}    
\end{small}

\begin{figure}[h]
\vspace{-0.1cm}
    \subfigtopskip=2pt
    \setlength{\belowdisplayskip}{2pt}
    \setlength{\abovedisplayskip}{-5pt}
    \subfigbottomskip=2pt
    \subfigcapskip=1pt
   \centering
    {\includegraphics[width=0.9\linewidth]{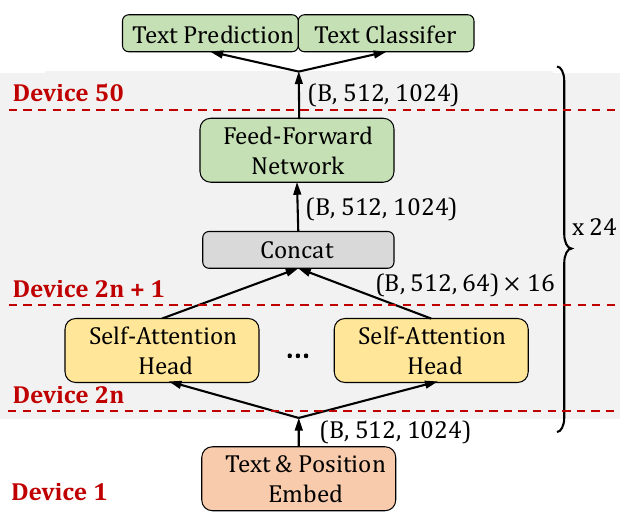}}
    \vspace{-0.1cm}
    \caption{Partitioned sub-DAGs of Bert-Large on 50 RTX 3080. There are 24 transformer layers, each of which is split into attention block and FFN block, }
    \label{fig:Bert-Large-SPLIT}
\vspace{-0.1cm}
\end{figure}

\textbf{Estimating system performance.}
Figure~\ref{fig:Bert-Large-SPLIT} shows an example of partitioning Bert-Large into multiple sub-DAGs and assigning them to 50 NVIDIA RTX 3080. And we also partition it into 4 NVIDIA H100 with sub-DAGs $1$, $2\sim25$, $26\sim49$ and $50$ in Figure~\ref{fig:Bert-Large-SPLIT} respectively. We coarsely estimate the computation time $\mathcal{C}_p$ based on FLOPs of sub-DAGs and TFLOPS (FP32 Tensor Core) of GPUs. Given the latency $\alpha$ and bandwidth $1/\beta$ between devices, we estimate the inference latency and cost of processing $n_b=512$ batches with pipeline by Equation~\ref{eq:time_latency} and ~\ref{eq:time_pipe}.

\begin{figure}[h!]
   \centering
    \subfigure{\includegraphics[width=0.23\textwidth]{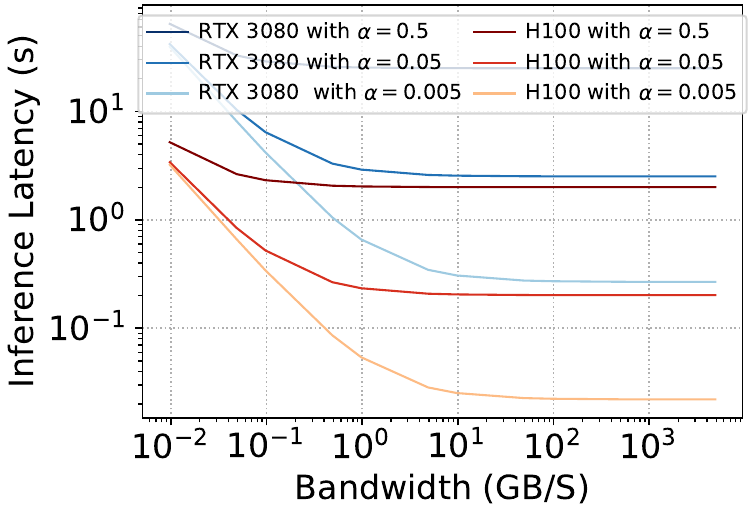}}
    \subfigure{\includegraphics[width=0.23\textwidth]{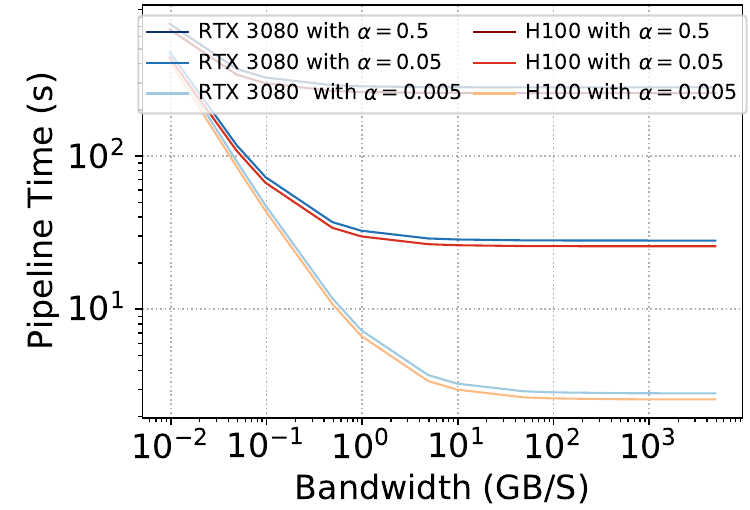}}
\vspace{-0.2cm}
\caption{\small System performance of Bert-Large with different communication bandwidth and latency.}
\label{fig:BertLargeSystem}
\end{figure}

\begin{figure}[h!]
   \centering
\vspace{-0.2cm}
    \subfigure{\includegraphics[width=0.23\textwidth]{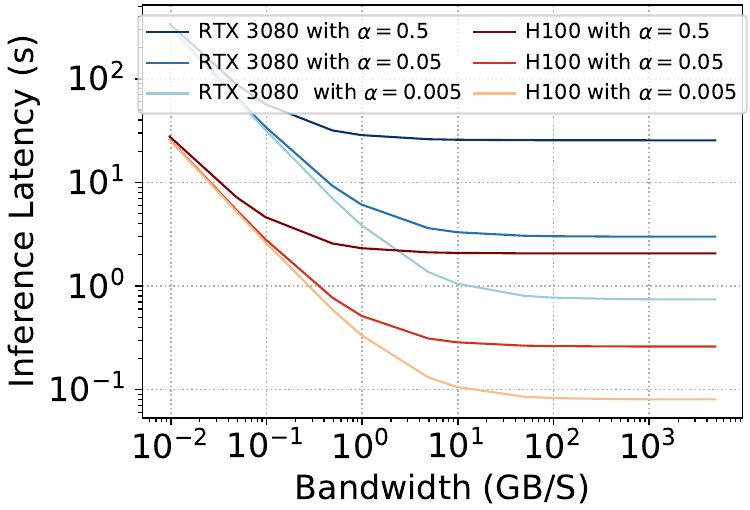}}
    \subfigure{\includegraphics[width=0.23\textwidth]{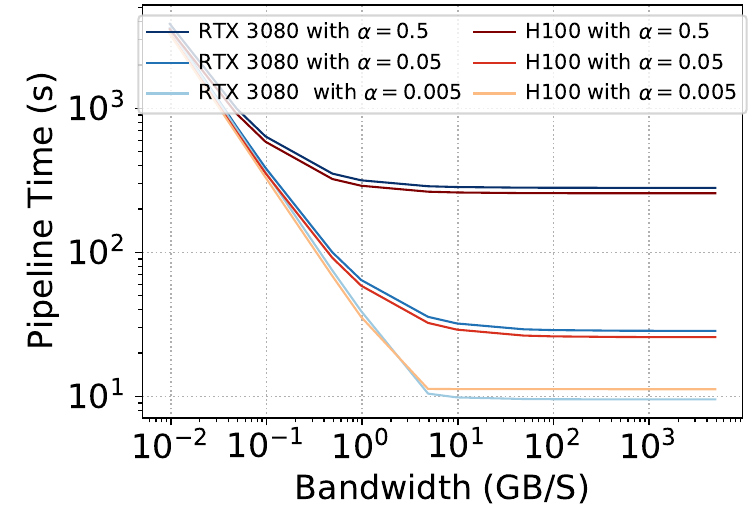}}
\vspace{-0.2cm}
\caption{\small System performance of GPT3 (24 layers with the hidden size of 4096).}
\label{fig:GPT3System}
\end{figure}

Figure~\ref{fig:BertLargeSystem} and~\ref{fig:GPT3System} shows the system performance of Bert and and GPT3. The latency time with 50$\times$ RTX 3080 is larger than 4$\times$ H100, due to more extra communication time from between more devices. But the throughput between them is similar, because the cost of pipeline is $(n_b - 1) \max_{p\in\mathcal{P}} (\mathcal{C}_p, \mathcal{R}_p)$, if $n_b$ is large. Thus, when we conduct continuous inference, large-scale RTX 3080 can achieve similar throughput with H100 but much lower prices. However, when training models, the pipeline may be cut off between different update iterations, and the activation must be cached, which severely limits the $n_p$. Future works may seek to design algorithms to solve this issue.


\section{Limitations}\label{sec:limitation}

\textbf{Fault tolerance.} In our current system design, the fault-tolerant method is handled in a relatively simple manner (Section~\ref{sec:broker}). The disconnection of compnodes interrupts DAG execution. Replacing disconnected peers with randomly sampled online peers can disrupt the load balance of scheduled tasks. Therefore, the costs of recovery, restart, and rescheduling need to be considered. Efficient fault tolerance schemes, including elastic training and swift and distributed checkpointing, will be explored and discussed in future work to improve the system's fault tolerance capabilities.

\textbf{Pipeline optimization.} To enhance the system efficiency, our load-balance scheduling (Section~\ref{sec:task-schedule}) provides an initial step towards reducing the bubble time of pipeline parallelism. However, it is still questioned how to efficiently execute pipelines on compnodes.

\section{Conclusion}\label{sec:conclusion}
In this work, we envision exploiting vast overlooked consumer-level GPUs for LLMs and privacy protection. We firstly identify the critical special challenges of this problem. Then, we proposed a system design as an initial step towards achieving this goal. We hope that this work will inspire ML system researchers to design and enhance decentralized training systems, enabling the utilization of the vast computing power offered by consumer-level GPUs. Furthermore, the methodologies developed in this system paves the way for training future giant LLMs that are potentially larger than GPT-4 with decentralized large-scale GPUs from different silos.

\section*{Acknowledgments}
This work was partially supported by National Natural Science Foundation of China under Grant No. 62272122, a Hong Kong RIF grant under Grant No. R6021-20, and a Hong Kong CRF grant under Grant No. C2004-21GF. 

\newpage

\bibliographystyle{ieeetr}
\bibliography{cite}

\end{document}